\newcommand{\PT}{{\rm P}_{\!\!\scriptscriptstyle\rm T}}
\newcommand{\ET}{{\rm E}_{\scriptscriptstyle\rm T}}
\newcommand{\MET}{\mbox{$\raisebox{.3ex}{$\not$}\ET$}}
\def\r#1{\ignorespaces $^{#1}$}
\font\eightit=cmti8

\documentstyle[preprint,aps,epsf]{revtex}
\tightenlines
\begin{document}
\draft
\title{Measurement of the Top Quark Mass}
\author{
\hfilneg
\begin{sloppypar}
\noindent
F.~Abe,\r {17} H.~Akimoto,\r {39}
A.~Akopian,\r {31} M.~G.~Albrow,\r 7 A.~Amadon,\r 5 S.~R.~Amendolia,\r {27} 
D.~Amidei,\r {20} J.~Antos,\r {33} S.~Aota,\r {37}
G.~Apollinari,\r {31}
% ...
\end{sloppypar}
\begin{center}
(CDF Collaboration)
\end{center}
\begin{center}
\r 1  {\eightit Argonne National Laboratory, Argonne, Illinois 60439} \\
\r 2  {\eightit Istituto Nazionale di Fisica Nucleare, University of Bologna,
I-40127 Bologna, Italy} \\
\r 3  {\eightit Brandeis University, Waltham, Massachusetts 02254} \\
\r 4  {\eightit University of California at Los Angeles, Los 
Angeles, California  90024} \\  
\r 5  {\eightit University of Chicago, Chicago, Illinois 60637} \\
\r 6  {\eightit Duke University, Durham, North Carolina  27708} \\
\r 7  {\eightit Fermi National Accelerator Laboratory, Batavia, Illinois 
60510} \\
\r 8  {\eightit University of Florida, Gainesville, FL  32611} \\
\r 9  {\eightit Laboratori Nazionali di Frascati, Istituto Nazionale di Fisica
               Nucleare, I-00044 Frascati, Italy} \\
\r {10} {\eightit University of Geneva, CH-1211 Geneva 4, Switzerland} \\
\r {11} {\eightit Harvard University, Cambridge, Massachusetts 02138} \\
\r {12} {\eightit Hiroshima University, Higashi-Hiroshima 724, Japan} \\
\r {13} {\eightit University of Illinois, Urbana, Illinois 61801} \\
\r {14} {\eightit Institute of Particle Physics, McGill University, Montreal 
H3A 2T8, and University of Toronto,\\ Toronto M5S 1A7, Canada} \\
\r {15} {\eightit The Johns Hopkins University, Baltimore, Maryland 21218} \\
\r {16} {\eightit Institut f\"{u}r Experimentelle Kernphysik, 
Universit\"{a}t Karlsruhe, 76128 Karlsruhe, Germany} \\
\r {17} {\eightit National Laboratory for High Energy Physics (KEK), Tsukuba, 
Ibaraki 305, Japan} \\
\r {18} {\eightit Ernest Orlando Lawrence Berkeley National Laboratory, 
Berkeley, California 94720} \\
\r {19} {\eightit Massachusetts Institute of Technology, Cambridge,
Massachusetts  02139} \\   
\r {20} {\eightit University of Michigan, Ann Arbor, Michigan 48109} \\
\r {21} {\eightit Michigan State University, East Lansing, Michigan  48824} \\
\r {22} {\eightit University of New Mexico, Albuquerque, New Mexico 87131} \\
\r {23} {\eightit The Ohio State University, Columbus, OH 43210} \\
\r {24} {\eightit Osaka City University, Osaka 588, Japan} \\
\r {25} {\eightit Universita di Padova, Istituto Nazionale di Fisica 
          Nucleare, Sezione di Padova, I-35131 Padova, Italy} \\
\r {26} {\eightit University of Pennsylvania, Philadelphia, 
        Pennsylvania 19104} \\   
\r {27} {\eightit Istituto Nazionale di Fisica Nucleare, University and Scuola
               Normale Superiore of Pisa, I-56100 Pisa, Italy} \\
\r {28} {\eightit University of Pittsburgh, Pittsburgh, Pennsylvania 15260} \\
\r {29} {\eightit Purdue University, West Lafayette, Indiana 47907} \\
\r {30} {\eightit University of Rochester, Rochester, New York 14627} \\
\r {31} {\eightit Rockefeller University, New York, New York 10021} \\
\r {32} {\eightit Rutgers University, Piscataway, New Jersey 08855} \\
\r {33} {\eightit Academia Sinica, Taipei, Taiwan 11530, Republic of China} \\
\r {34} {\eightit Texas A\&M University, College Station, Texas 77843} \\
\r {35} {\eightit Texas Tech University, Lubbock, Texas 79409} \\
\r {36} {\eightit Istituto Nazionale di Fisica Nucleare, University of Trieste/
Udine, Italy} \\
\r {37} {\eightit University of Tsukuba, Tsukuba, Ibaraki 315, Japan} \\
\r {38} {\eightit Tufts University, Medford, Massachusetts 02155} \\
\r {39} {\eightit Waseda University, Tokyo 169, Japan} \\
\r {40} {\eightit University of Wisconsin, Madison, Wisconsin 53706} \\
\r {41} {\eightit Yale University, New Haven, Connecticut 06520} \\
\end{center}
}
\date{September 30, 1997}
\maketitle
\begin{abstract}
  We present a measurement of the top quark mass using a sample of 
  $t\bar{t}$ decays into an electron or a muon, a neutrino, and four jets.
  The data were collected in $p\bar{p}$ collisions at 
  $\sqrt{s}=1.8$~TeV with the Collider Detector at Fermilab
  and correspond to an integrated luminosity of 109 pb$^{-1}$.
  We measure the top quark mass to be 
  175.9 $\pm$ 4.8(stat.) $\pm$ 4.9(syst.) GeV/$c^{2}$. 
\end{abstract}
\pacs{14.65.Ha, 13.85.Qk, 13.85.Ni}

The top quark mass is a fundamental parameter of the standard model and is 
needed for extracting other parameters from precision electroweak measurements.
The first direct measurement of its value was made by 
CDF~\cite{Evidence_papers} and was based on 19~pb$^{-1}$ of data.  
Updated measurements were reported by both the CDF and D\O\ collaborations 
using significantly more 
data~\cite{CDF_discovery,D0_discovery,D0_mass_PRL,CDF_hadronic,CDF_dilepton}.
In this paper we present a new measurement of the top quark mass with 
greatly improved precision, using our entire data sample from the 1992--1995 
runs, which corresponds to a total integrated luminosity of 
$109 \pm 7$~pb$^{-1}$~\cite{Xsection_PRL}.  This new measurement supersedes
the results reported in~\cite{Evidence_papers,CDF_discovery}. 

Within the standard model, the top quark decays more than 99\% of the time
into $Wb$.  The $W$ boson can then decay to a quark-antiquark or 
lepton-neutrino pair.  The measurement presented here uses events with a 
$t\bar{t}$ pair decaying in the ``lepton+jets'' channel.  This channel is 
characterized by a single high-$\PT$~\cite{Evidence_papers} lepton 
(electron or muon) and 
missing transverse energy from a $W\rightarrow\ell\nu$ decay, plus 
several jets coming from a hadronically decaying $W$ boson and from the $b$ 
quarks from the top quark decays.  Jets formed by the fragmentation of $b$ 
quarks can be identified (``tagged'') either by reconstructing secondary 
vertices from $b$ hadron decays with the silicon vertex detector (SVX tagging), 
or by finding additional leptons from semileptonic $b$ decays (SLT tagging).  
The SVX and SLT tagging algorithms are described in Ref.~\cite{CDF_discovery}.  

To be used for the mass measurement, events must contain a single
isolated electron (muon) with $\ET\;(\PT) > 20$ GeV (GeV/$c$) in the 
central region of the detector ($|\eta| < 1$) and missing 
transverse energy, $\MET\geq 20$ GeV, indicating the presence 
of a neutrino.  At least four jets are required in each event, three
of which must have an observed $\ET\geq 15$ GeV and $|\eta| \leq 2$.  
In order to increase the acceptance, we relax the requirements 
on the fourth jet to be $\ET\geq 8$ GeV and $|\eta|\leq 2.4$, provided one
of the four leading jets is tagged by the SVX or SLT algorithms.  
SVX tags are only
allowed on jets with observed $\ET\geq 15$ GeV, while SLT tags are allowed 
on jets with $\ET\geq 8$ GeV.  If no such tag is present,
the fourth jet must satisfy the same $\ET$ and $\eta$ requirements as the 
first three.  All jets in this analysis are formed as clusters of calorimeter 
towers within cones of fixed radius 
$\Delta R\equiv\sqrt{\Delta\eta^{2}+\Delta\phi^{2}}=0.4$~\cite{Jet_clustering}. 
The above selection defines our mass sample, which contains 83 events.

Measurement of the top quark mass begins by fitting each event
in the sample to the hypothesis of $t\bar{t}$ production followed by
decay in the lepton+jets channel:
\begin{eqnarray*}
p\,\bar{p} & \longrightarrow & t\,\bar{t} + X \\
           &                 & 
\begin{array}{l} 
   t \longrightarrow W^{+} \, b \longrightarrow \ell^{+}\, \nu\, b \\
   \bar{t} \longrightarrow W^{-} \, \bar{b} \longrightarrow 
   q\,\bar{q}^{\,\prime}\,\bar{b}
\end{array}
\;\;\;{\rm or}\;\;\;
\begin{array}{l}
   q\,\bar{q}^{\,\prime}\,b \\
   \ell^{-}\,\bar{\nu}\, \bar{b}\, .
\end{array}
\end{eqnarray*}
The 3-momenta of the lepton and the $b$, $\bar{b}$, $q$ and $\bar{q}^{\,\prime}$
quarks are measured from the observed lepton and four leading jets in the event;
the mass of the $b$ is set to 5 GeV/$c^{2}$, that of $q$ and 
$\bar{q}^{\,\prime}$ to 0.5 GeV/$c^{2}$.  The neutrino mass is assumed to be 
zero and its momentum is not measured, thereby yielding three unknowns.
The two transverse momentum components of $X$ are measured from the extra
jets in the event and the energy that is detected but not collected in jet
or electron clusters.  Five constraints are applied:  the transverse momentum
components of the entire $t\bar{t} + X$ system must be zero, the invariant 
masses of the lepton-neutrino and $q$-$\bar{q}^{\,\prime}$ pairs must each 
equal the $W$ boson mass, and the mass of the top quark must equal that of the 
antitop quark.  The problem therefore has two extra constraints and is solved
by a standard $\chi^{2}$-minimization technique.  The output of each event fit 
is a reconstructed top mass $M_{\rm rec}$ and a $\chi^{2}$ value quantifying
how well the event is described by the $t\bar{t}$ hypothesis.

Electron energies and muon momenta entering the fit are measured with the 
calorimeter and tracking chambers, respectively~\cite{W_mass_1A}.  Jet 
energies are corrected for losses in cracks between detector components, 
absolute energy scale, contributions from the underlying event and multiple 
interactions, and losses outside the clustering cone.  These corrections are 
determined from a combination of Monte Carlo simulations and 
data~\cite{Jet-energy-scale}.  The four leading jets in a $t\bar{t}$ candidate 
event undergo an additional energy correction that depends on the type of 
parton they are assigned to in the fit: a light quark, a hadronically decaying 
$b$ quark, or a $b$ quark that decayed semileptonically~\cite{Evidence_papers}. 
This parton-specific correction was derived from a study of $t\bar{t}$ events 
generated with the {\sc herwig} Monte Carlo 
program~\cite{Herwig,Structure-functions}.

There are twelve distinct ways of assigning the four leading jets to the 
four partons $b$, $\bar{b}$, $q$, and $\bar{q}^{\,\prime}$.  
In addition, there is a quadratic ambiguity in the determination of the 
longitudinal component of the neutrino momentum.  This yields up to 
twenty-four different configurations for reconstructing 
an event according to the $t\bar{t}$ hypothesis.  We require that SVX or 
SLT-tagged jets be assigned to $b$-partons and choose the configuration
with lowest $\chi^{2}$.  Events with $\chi^{2} > 10$ are rejected.
In the mass sample, 76 out of 83 events remain after this cut.
When all parton-jet assignments are correctly made, the resolution of
the reconstructed mass is 13 GeV/$c^{2}$ for a top mass of 175 GeV/$c^{2}$.

A maximum-likelihood method is used to extract a top mass measurement from
a sample of events which have been reconstructed according to the 
$t\bar{t}$ hypothesis.  An essential ingredient of the likelihood function
is the probability density $f_{s}(M_{\rm rec}|M_{\rm top})$ to reconstruct a
mass $M_{\rm rec}$ from a $t\bar{t}$ event if the true top mass is 
$M_{\rm top}$.  In past publications~\cite{Evidence_papers,CDF_discovery} we 
estimated $f_{s}$ for a discrete set of $M_{\rm top}$ values by smoothing 
histograms of $M_{\rm rec}$ for events from a {\sc herwig} Monte Carlo 
calculation. In the present analysis we parameterize $f_{s}$ as a smooth 
function of both $M_{\rm rec}$ and $M_{\rm top}$~\cite{Bettelli}.  
This new approach yields a consistent,
$M_{\rm top}$-dependent way of dealing with low statistics in the tails of
the $M_{\rm rec}$ histograms and produces a continuous likelihood
shape from which the top mass and its uncertainty can be extracted.
The probability density $f_{b}(M_{\rm rec})$ for reconstructing a
mass $M_{\rm rec}$ from a background event is obtained by fitting a smooth
function to a mass distribution generated with the {\sc vecbos}~\cite{Vecbos} 
W+jets Monte Carlo program.

The likelihood function is the product of three factors:
\begin{equation}
{\cal L} = {\cal L}_{shape}\times {\cal L}_{backgr} \times {\cal L}_{param}\, ,
\label{Likelihood}
\end{equation}
where ${\cal L}_{shape}$ represents the joint probability density for
a sample of $N$ reconstructed masses $M_{i}$ to be drawn from a population
with a background fraction $x_{b}$:
\begin{displaymath}
{\cal L}_{shape} = \prod_{i\,=\,1}^{N}\left[
                       (1-x_{b})\,f_{s}(M_{i}| M_{\rm top}) + 
                       x_{b}\,f_{b}(M_{i})\right]\, .
\end{displaymath}
The fraction $x_{b}$ is constrained by an independent measurement that is 
summarized by the background likelihood ${\cal L}_{backgr}$.  
The function ${\cal L}_{param}$ allows the parameterizations of $f_{s}$ and 
$f_{b}$ to vary within the uncertainties returned by the fits to the 
{\sc herwig} and {\sc vecbos} histograms of $M_{\rm rec}$.
By including ${\cal L}_{param}$ in the likelihood definition, the 
uncertainty due to the finite statistics of these histograms is incorporated 
into the statistical uncertainty on the measured top mass.
The likelihood ${\cal L}$ is maximized with respect to $M_{\rm top}$, 
$x_{b}$, and the parameters that define the shapes of $f_{s}$ and $f_{b}$.

The precision of the top quark mass measurement is expected to increase with 
the number of observed events, the signal-over-background ratio, and the 
narrowness of the reconstructed-mass distribution.  
These characteristics vary significantly between samples with different 
$b$ tagging requirements.  
Therefore, to make optimal use of all the available information,
we partition the mass sample into non-overlapping subsamples, define 
subsample likelihoods according to eq.~(\ref{Likelihood}), and maximize
the product of these likelihoods to determine the top mass and its 
uncertainty~\cite{Tollefson}.
The use of non-overlapping subsamples ensures that the corresponding
likelihoods are statistically uncorrelated.
Monte Carlo studies show that an optimum partition is 
made up of four subsamples: events with a single SVX tag, events with two 
SVX tags, events with an SLT tag but no SVX tag, and events with no tag
but with the tighter kinematic requirement of four jets with 
$\ET\geq 15$ GeV and $|\eta|\leq 2$.  

The calculation of the expected background content of each subsample starts
from the background calculation performed on the $W+\geq3$-jet sample
for the $t\bar{t}$ cross section measurement~\cite{Xsection_PRL}.  
The extrapolation to the mass subsamples takes into account
the additional requirement of a fourth jet, the $\chi^{2}$ cut on 
event reconstruction, and the fact that SVX and SLT tags are only counted
if they are on one of the four leading jets.  The efficiencies of these 
requirements are determined from Monte Carlo studies.  They are used together 
with background rates and tagging efficiencies from the cross section analysis 
to predict the total number of events in each mass subsample as a function of
the unknown numbers of $t\bar{t}$ and $W$+jet events in the combined sample.  
These unknowns are estimated by maximizing a multinomial likelihood that
constrains the predicted subsample sizes to the observed ones.  
This procedure generates the expected background fractions shown 
in Table~\ref{Top_masses} and the background likelihood ${\cal L}_{backgr}$
used in eq.~(\ref{Likelihood}).

Approximately 67\% of the background in the entire mass sample comes from
$W$+jet events.  Another 20\% consists of multijet events where a jet is 
misidentified as a lepton and $b\bar{b}$ events with a $b$ hadron decaying 
semileptonically.  The remaining 13\% is made up of $Z$+jet events where the
$Z$-boson decays leptonically, events with a $WW$, $WZ$ or $ZZ$ diboson, 
and single-top production.
We have compared the reconstructed-mass distributions in {\sc vecbos} and 
data for three event selections that are expected to be depleted in $t\bar{t}$ 
events~\cite{Eddy}.  These selections are slight variations of the 
mass sample selection.  The first one requires that the primary lepton be 
an electron with a pseudo-rapidity in the range $1.1\leq |\eta|\leq 2.4$ 
instead of $|\eta|\leq 1$, and yields 26 data events.  The second one requires 
at least four jets with $\ET\geq 8$ GeV and $|\eta|\leq 2.4$, but no more than 
two jets with $\ET\geq 15$ GeV and $|\eta|\leq 2$.  This results in 243 data
events.  The third selection requires events with a 
non-isolated primary lepton and yields 164 data events.  
In all three cases, a Kolmogorov-Smirnov test applied to the comparison of
{\sc vecbos} and data yields a confidence level of at least 30\%.
We therefore use the {\sc vecbos} calculation to determine the shape 
of $f_{b}$ for the likelihood function.

\begin{figure}
\epsffile[18 122 527 648]{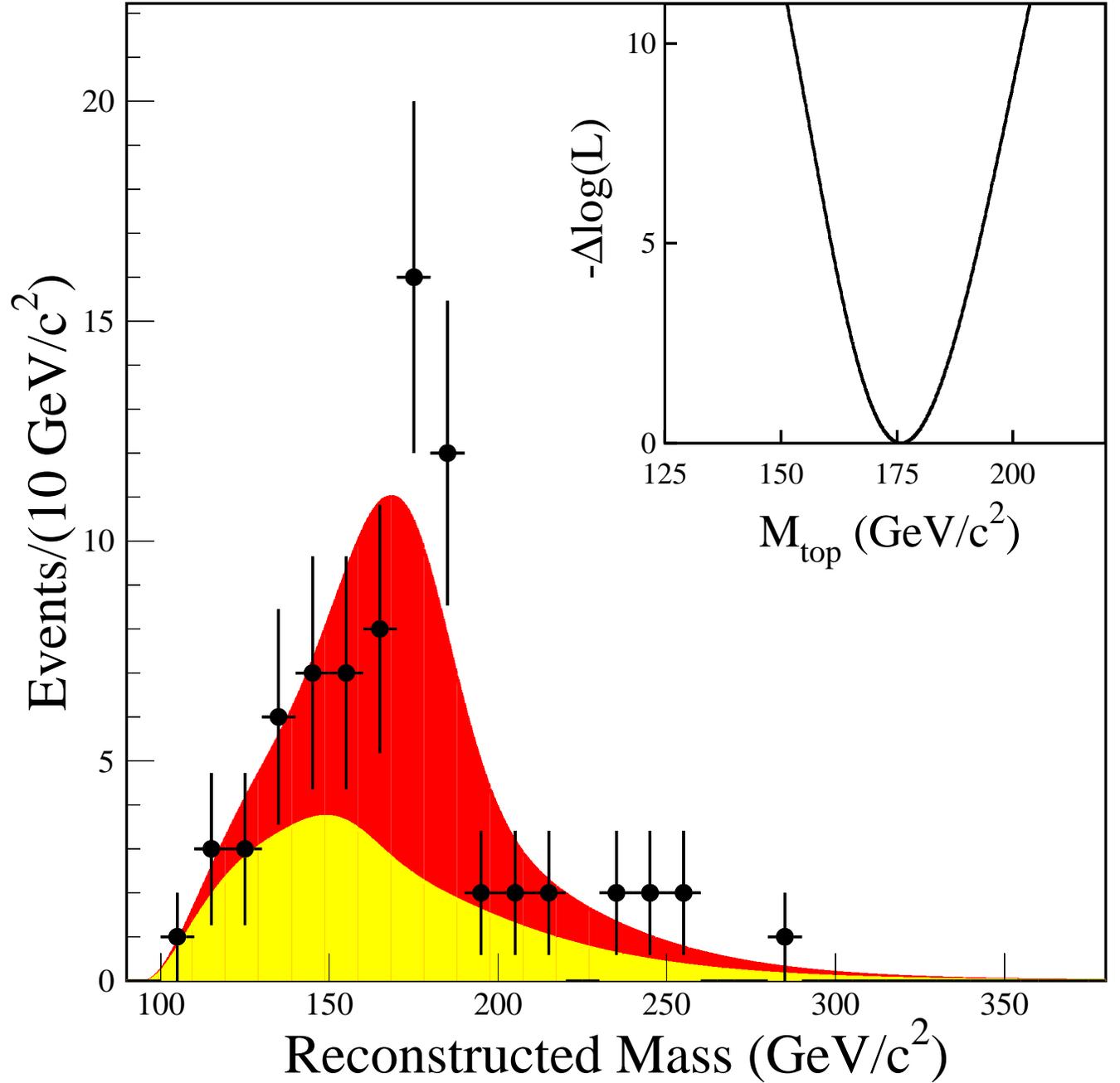}
\caption{Reconstructed-mass distribution of the four mass
subsamples combined.  The data (points) are compared with the 
result of the combined fit (dark shading) and with the background
component of the fit (light shading).
The inset shows the variation of the combined negative log-likelihood
with $M_{\rm top}$. 
\label{Mfit_combined}}
\end{figure}

\begin{figure}
\epsffile[18 122 527 675]{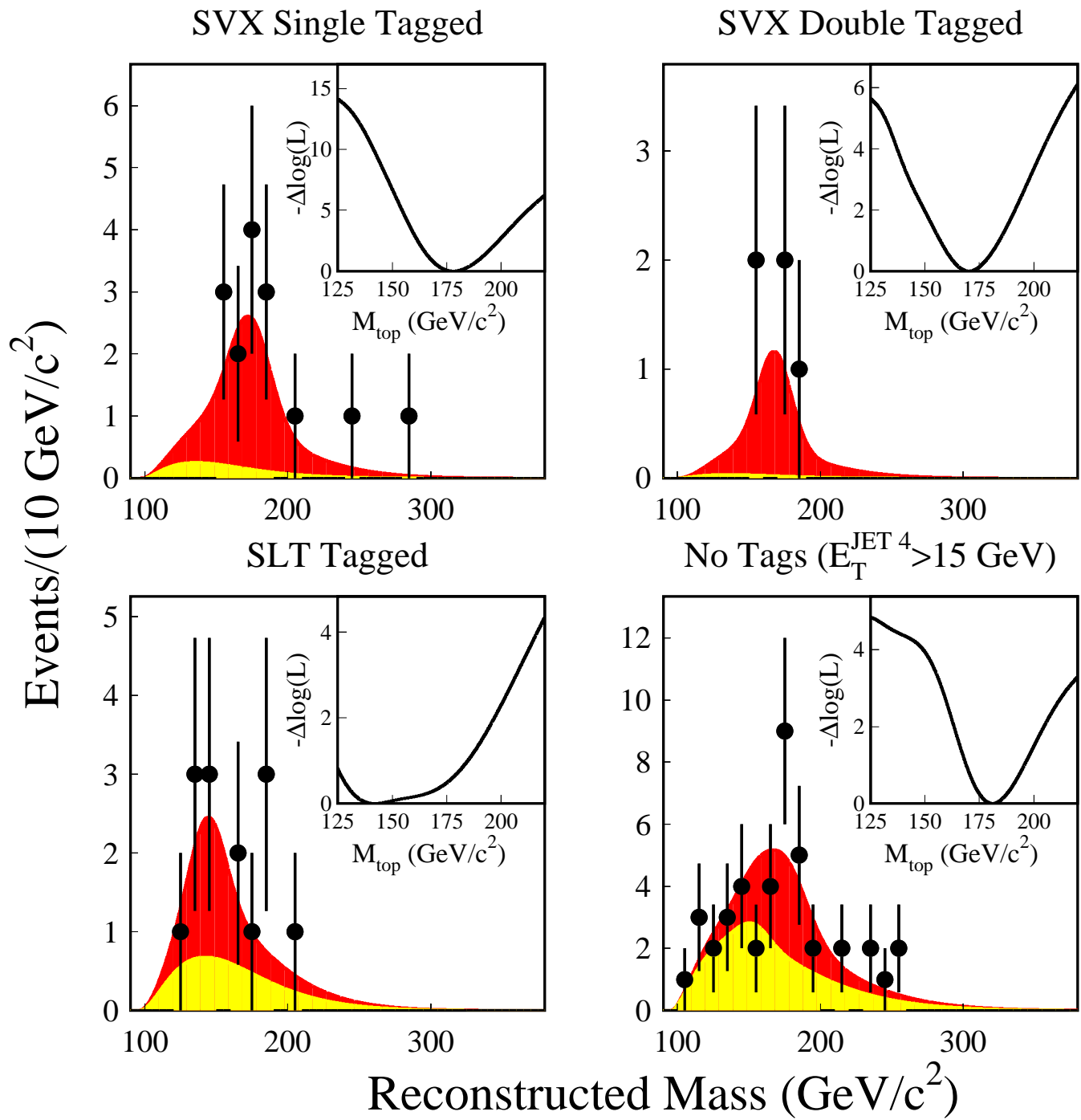}
\caption{Reconstructed-mass distributions in each of the four mass
subsamples.  Each plot shows the data (points), the result of the combined fit
to top+background (dark shading), and the background component of the fit 
(light shading).  The insets show the variation of the negative log-likelihoods 
with $M_{\rm top}$ for the separate subsample fits.
\label{Mfit_separate}}
\end{figure}

The reconstructed-mass distribution of the sum of the four subsamples is 
plotted in Figure~\ref{Mfit_combined}.  The inset shows the shape of the 
corresponding sum of negative log-likelihoods as a function of top mass.  
From this we measure $M_{\rm top} = 175.9\pm 4.8$ GeV/$c^{2}$, where 
the uncertainty corresponds to a half-unit change in the negative 
log-likelihood with respect to its minimum.  Monte Carlo studies on mass 
samples similar to ours yield an 11\% probability for obtaining a statistical 
uncertainty of this size or smaller.  The background fractions $x_{b}$
returned by the fit agree with the $x_{b}^{0}$ numbers listed in 
Table~\ref{Top_masses}.  To judge the goodness of the fit of the combined
$M_{\rm rec}$ distribution, we performed a Kolmogorov-Smirnov test and 
obtained a confidence level of 64\%.  The reconstructed-mass distribution 
in each of the four subsamples is compared to the result of the combined fit 
in Figure~\ref{Mfit_separate}.  The insets show the results of likelihood fits 
performed separately in each of the four subsamples.  The mass measurements 
obtained from these fits are consistent with each other, as shown in 
Table~\ref{Top_masses}.

We list the systematic uncertainties in Table~\ref{Systematics}.
The largest one comes from the jet energy measurement.  Each of the jet
energy corrections described earlier carries with it a separate,
energy-dependent uncertainty~\cite{Jet-energy-scale}.  
Recent studies of soft gluon radiation outside the jet clustering cone have
reduced the uncertainty from this source to 2.5\% for a jet with observed
$\ET > 40$ GeV.  For an observed jet $\ET$ of 40 GeV, 
the total uncertainty on the corrected $\ET$ varies between 3.4 and 5.6\% 
depending on the proximity of the jet to cracks between detector components.
We have checked the jet correction procedure and the evaluation of
the jet energy scale uncertainty with events containing a leptonically
decaying $Z$ boson and one jet.  A study of how the transverse momentum of the
jet balances that of the $Z$ decay products finds that the observed ratio
of $[\PT(Z) - \PT({\rm jet})]/\PT(Z)$ differs by
3.2 $\pm$ 1.5(stat.) $\pm$ 4.1(syst.) \% from Monte Carlo simulations.
The 4.1\% systematic uncertainty is due to the jet energy scale only.  
Since the difference is consistent with zero, this study independently 
confirms the soundness of our estimate of the jet energy scale uncertainty.
A further confirmation was obtained by measuring the mass of the $W$ boson from
its hadronic decay modes, using a sample of $t\bar{t}$ candidate events in the
lepton+jets channel.  This measurement yields
$77.2\pm 3.5 {\rm (stat.)}\pm 2.9 {\rm (syst.)}$ GeV/$c^{2}$~\cite{Wjj_PRL}.

The second largest systematic uncertainty is due to high transverse 
momentum gluons that are radiated from the initial or final state of 
a $t\bar{t}$ event and sometimes take the place of a $t\bar{t}$ decay
product among the four leading jets.  This uncertainty was determined 
with the {\sc pythia} Monte Carlo calculation~\cite{Pythia} by separately
studying the effect of extra jets coming from initial and final state 
radiation.

The uncertainty in the modeling of the background mass distribution was 
estimated by varying the $Q^{2}$ scale in {\sc vecbos}.
Additional sources of uncertainty include the kinematical bias introduced by 
$b$ tagging and the choice of parton distribution functions 
({\sc cteq4l}~\cite{CTEQ4} vs. {\sc mrsd0$^{\prime}$}).  
The sum in quadrature of all the systematic 
uncertainties is 4.9 GeV/$c^{2}$.  We have investigated the effect of using 
Monte Carlo calculations other than {\sc herwig} to model $t\bar{t}$ events.  
Whereas {\sc pythia} yields the same measured mass, {\sc isajet}~\cite{Isajet} 
leads to a $+1.5$ GeV/$c^{2}$ shift.  We do not include this as a separate 
uncertainty since the main difference between these calculations, namely the
modeling of gluon radiation and jet fragmentation, is already
accounted for in our analysis of other systematic uncertainties.

In summary, we have measured the top quark mass to be
175.9 $\pm$ 4.8(stat.) $\pm$ 4.9(syst.) GeV/$c^{2}$.  This is the most
precise determination of the top mass to date.  
A new technique for optimizing the use of the information provided by 
the tagging algorithms has resulted in a smaller statistical uncertainty,
and a better understanding of the jet energy scale has led to a reduced
systematic uncertainty.  In addition, the probability densities for 
reconstructed masses are now fully parameterized, which simplifies
the likelihood analysis and the treatment of the finite statistics of the
Monte Carlo event samples.

We thank the Fermilab staff and the technical staffs of the participating
institutions for their vital contributions. This work is supported
by the U.S. Department of Energy and the National Science Foundation, the
Natural Sciences and Engineering Research Council of Canada, the
Istituto Nazionale di Fisica Nucleare of Italy, the Ministry of 
Education, Science and Culture of Japan, the National Science Council of
the Republic of China, and the A.P. Sloan Foundation.

\begin{table}
\caption{Subsamples of $W+\geq 4$-jet events that are used for the top quark 
mass measurement.  For each subsample, the number of observed events $N_{obs}$,
the expected background fraction $x_{b}^{0}$, and the measured top mass 
$M_{\rm top}$ are shown.
Uncertainties on the measured top mass are statistical only.}
\begin{tabular}{cccc}
                          &         &  $x_{b}^{0}$ & Measured $M_{\rm top}$ \\
Subsample                 &$N_{obs}$&     (\%)     &   (GeV/$c^{2}$)    \\
\tableline
SVX double tag                  & 5  & $5\pm 3$   & $170.1\pm 9.3$    \\
SVX single tag                  & 15 & $13\pm 5$  & $178.0\pm 7.9$    \\
SLT tag (no SVX)                & 14 & $40\pm 9$  & $142^{+33}_{-14}$ \\
No tag ($\ET(j_{4})\geq 15$ GeV)& 42 & $56\pm 15$ & $181.0\pm 9.0$    \\
\end{tabular}
\label{Top_masses}
\end{table}

\begin{table}
\caption{List of systematic uncertainties on the final top quark mass 
measurement.}
\begin{tabular}{cc}
Source                            & Value (GeV/$c^{2}$)  \\
\tableline
Jet energy measurement            &      4.4             \\
Initial and final state radiation &      1.8             \\
Shape of background spectrum      &      1.3             \\
$b$ tag bias                      &      0.4             \\
Parton distribution functions     &      0.3             \\
\tableline
Total                             &      4.9             \\
\end{tabular}
\label{Systematics}
\end{table}

\end{document}